# Coupling between electrons and charge density wave fluctuation and its possible role in superconductivity

Yeonghoon Lee[1,2], Yeahan Sur[3], Sunghun Kim[1,4], Jaehun Cha[1], Jounghoon Hyun[1], Chan-young Lim[1], Makoto Hashimoto[5], Donghui Lu[5], Younsik Kim[6,7], Soonsang Huh[6,7], Changyoung Kim[6,7], Shinichiro Ideta[8,9], Kiyohisa Tanaka[8], Kee Hoon Kim[3,10], Yeongkwan Kim[1]*

[1]Department of Physics, Korea Advanced Institute of Science and Technology, Daejeon 34141, Republic of Korea.

[2]Quantum Technology Institute, Korea Research Institute of Standards and Science, Daejeon 34113, Republic of Korea.

[3]Center for Novel States of Complex Materials Research, Department of Physics and Astronomy, Seoul National University, Seoul 08826, Republic of Korea.

[4]Department of Physics, Ajou University, Suwon 16499, Republic of Korea.

[5]Stanford Synchrotron Radiation Light Source, Stanford Linear Accelerator Center, Menlo Park, CA, 94025, USA.

[6]Center for Correlated Electron Systems, Institute for Basic Science, Seoul 08826, Republic of Korea.

[7]Department of Physics and Astronomy, Seoul National University, Seoul 08826, Republic of Korea.

[8]Ultra Violet Synchrotron Orbital Radiation, Institute for Molecular Science, Myodaiji, Okazaki 444-8585, Japan.

[9]Hiroshima Synchrotron Radiation Center, Hiroshima University, Higashi-Hiroshima 739-0046, Japan.

[10]Institute of Applied Physics, Seoul National University, Seoul 08826, Republic of Korea.

*e-mail: yeongkwan@kaist.ac.kr

**In most of charge density wave (CDW) systems of different material classes, ranging from traditional correlated systems in low-dimension to recent topological systems with Kagome lattice, superconductivity emerges when the system is driven toward the**

**quantum critical point (QCP) of CDW via external parameters of doping and pressure. Despite this rather universal trend, the essential hinge between CDW and superconductivity has not been established yet. Here, the evidence of coupling between electron and CDW fluctuation is reported, based on a temperature- and intercalation-dependent kink in the angle-resolved photoemission spectra of $2H$-Pd$_x$TaSe$_2$. Kinks are observed only when the system is in the CDW phase, regardless of whether a long- or short-range order is established. Notably, the coupling strength is enhanced upon long-range CDW suppression, albeit the coupling energy scale is reduced. Interestingly, estimation of the superconducting critical temperature by incorporating the observed coupling characteristics into McMillan's equation yields result closely resembling the known values of the superconducting dome. Our results thus highlight a compelling possibility that this new coupling mediates Cooper pairs, which provides new insights on the competing relationship not only for CDW, but also for other competing orders.**

1. Introduction

It is widely documented that superconductivity emerges near the quantum critical point (QCP) of symmetry-broken phases[1-9]. This competing behavior, or presence of a competing order, has led to notable speculation that quantum fluctuations of the order could play a role in superconductivity formation, thereby pairing electrons into Cooper pairs. For instance, spin fluctuation has been considered a pairing mediator in cuprate, iron-based, and heavy fermion superconductors where spin ordering competes with superconductivity[10–16]. A CDW, an ordering of itinerant charge carriers, also exhibits competing behavior with the superconductivity[1-9]. Therefore, similar to the other cases, it is natural to speculate that a CDW-associated low-energy excitation could pair the electrons and induce the superconductivity, which has not been sufficiently visited.

The first step in unveiling the competing behavior between CDW and superconductivity is to investigate whether electrons indeed couple with CDW-originated excitations (Figure 1a). Candidates include the zone-folded phonon (ZFP) due to a reduced translational symmetry and collective CDW excitation modes of amplitude oscillation (amplitudon) and phase alternation (phason). Once an electron couples with CDW-originated excitations, specifically with the amplitudon (e-amp coupling), this should be manifested in the low-energy electronic structure. Normally, when an electron couples with a bosonic mode such as the phonon (e-ph coupling), abrupt renormalization of electron band dispersion so called kink occurs, which determines the energy scale of the coupled mode and coupling strength[17-19]. In contrast to e-ph coupling, the kink attributable to CDW-originated excitations should be temperature-dependent, especially the energy scale, since the energy scale of amplitudon is proportional to the CDW order parameter, which reduces to zero at the CDW transition temperature $T_{CDW}$ (Figure 1b).

In this work, we are able to capture such temperature-dependent kinks, the smoking-gun evidence of e-amp coupling, in the low energy electronic structure of the representative CDW system, $2H$-TaSe$_2$. Further, the systematic investigation across the phase diagram spanned by Pd intercalation reveals that, when the system approaches the QCP of CDW, e-amp coupling is strengthened, which successfully explains the superconductivity enhancement near the QCP. Our work thus implies a plausible role of this coupling in the superconductivity, bridging CDW and superconductivity.

2. Results

2.1. Proper Momentum Positions for Investigating Kink

To explicitly search for these temperature-dependent kinks, alternative temperature-dependent band renormalizations associated with the CDW transition should be avoided. Upon cooling from the normal phase, TaSe$_2$ first enters the incommensurate CDW (ICCDW) phase at $T_{ICC}$, and the transition into commensurate CDW (CCDW) occurs at a lower temperature, i.e., $T_{CC}$. Between these two transitions, there is a region where CCDW and ICCDW coexist, referred to as the coexisting phase[20, 21]. This series of transitions leaves a footprint in the electronic structure, such as the pseudogap, band folding, and CDW gap[22–24]. To avoid those renormalizations, especially the complicated gap opening at the Fermi level, possible kinks in the band along the Γ-K and M-K lines were investigated (Figure 1c).

## 2.2. Electron-Amplitudon Kink via Temperature-Dependent Measurements

Figure 2 shows band dispersion along the Γ-K and M-K lines at different temperatures. At low temperatures (Figure 2a and 2b), kinks are relatively clearly observed in both the Γ-K and M-K lines, respectively. The peak positions obtained via momentum distribution curve (MDC) fitting (see Supplementary Information) reveal the presence of kinks at two different binding energies in both high-symmetry lines (Figure 2c and 2d). The low-energy dispersion clearly deviates from the estimated bare band dispersion at the two binding energies (the dashed lines, see Supplementary Information for estimation). Notably, in both lines, low-energy kinks disappear at a high temperature of 130 K, which is above $T_{CC}$ and $T_{ICC}$.

To characterize these kinks, the real part of self-energy (ReΣ) is extracted as shown in Figure 2e and 2f for the Γ-K and M-K lines, respectively, by subtracting the bare band dispersion from fitted peak positions[17]. The energy points of sudden slope change determine the kink energy scales, 9 meV for low-energy kink and 22 meV for high-energy kink, as identified by the black vertical lines (see Supplementary information for validity of extracted ReΣ and determination of kink energy scales). The energy scales of the observed kinks in both the Γ-K and M-K lines are almost identical. In regard to high-energy kinks, the kink energy scale is much larger than the known collective excitations and does not freeze to zero (Figure 2g), suggesting that high-energy kinks originate from either the ordinary phonon or ZFP. In sharp contrast, with increasing temperature, the lower-energy kinks exhibit softening to an energy scale ranging from 4–6 meV before finally disappearing (Figure 2h). The thermal effect – the broadening of electron spectral weight itself and the weakening of coupling strength, could cause the disappearance of the kink with low energy scale at high temperatures. However, the thermal effect does not provoke the change in the kink energy (see Figure S8). Thus, the observed softening requires another origin, which is not expected for the phonon but is adequate for the CDW amplitudon[25]. Indeed, the observed energy scale and associated softening behavior closely match those of the $A_{1g}$ amplitudon measured through Raman spectroscopy[26–28], suggesting the e-amp coupling origin for the lower-energy kinks. Further, no other candidates fit into the observations; i) there is no optical phonon with energy less than 15 meV[26-30], ii) the acoustic phonons cannot explain the isotropic nature of low-energy kink (see Figure S10 and related texts), iii) sudden drop of the coupling strength above $T_{CDW}$ does not fit to phonons (Figure 2i), iv) the CDW is the only order which generates the collective excitation mode other than phonons, v) the CDW phason energy does not severely change while increasing temperature[26–28]. Therefore, we conclude these low-energy (high-energy) kinks as e-amp (e-ph) kinks.

Fitting the temperature dependence of e-amp kinks with the formula $\omega(T) = \omega(0)(1 - T/T_{CDW})^{0.19}$ yields a $T_{CDW}$ value of 102 K (the gray curve in Figure 2h). The exponent value of 0.19 was retrieved from the previous Raman study[26]. It should be noted that, instead of $T_{CC}$, the fitted $T_{CDW}$ value reaches 107 K, where the coexisting phase ends, suggesting that e-amp kinks persist up to the coexisting phase, where CCDW is expected to establish a short-range order. This somewhat unexpected behavior was clarified further through systematic investigation of the kinks across the phase diagram derived via Pd intercalation, as shown in Figure 3. The filled markers in Figure 3a indicate the points where e-amp kinks are observed. The empty markers indicate the absence of kinks. This suggests that e-amp kinks persist in the Pd-intercalated system up to the coexisting phase at all intercalation levels, even at high levels where the long-range CCDW no longer remains stable at any temperature. Moreover, the suppression of the e-amp kink energy is repeated with increasing temperature at each level (see Supplementary Information). This strongly suggests that the amplitudon persists up to the coexisting phase instead of $T_{CC}$, which will be discussed later.

### 2.3. Evolution of Kink Energy and Coupling Strength Upon Pd Intercalation

To trace the evolution of kinks upon Pd intercalation, the lowest-temperature dispersions are compared between the different intercalation levels, as shown in Figure 3b and 3c. The spectral weight tends to broaden, presumably due to the induced inhomogeneity. Nevertheless, the precise band dispersions obtained through MDC fitting (Figure 3d and 3e) clearly illustrate both e-amp and e-ph kinks, respectively, persisting up to the highest level in $Pd_{0.12}TaSe_2$. Since Pd intercalation is expected to modify the bare band dispersion, further analysis was carried out with ReΣ, as shown in Figure 4a and 4b for both the Γ-K and M-K lines, respectively. The extracted ReΣ indicates that the e-ph kink energy remains almost intact against Pd intercalation (Figure 4c). This seems reasonable since Pd intercalation between $TaSe_2$ layers should hardly affect the phonon[2]. Meanwhile, e-amp kinks exhibit progressive softening upon Pd intercalation (Figure 4d). This is the expected trend since the magnitude of the CDW order parameter decreases upon Pd intercalation, as evidenced by the reduced $T_{CC}$, which eventually disappears at a high intercalation level. Consistently, amplitudon softening upon CDW phase suppression was observed in a similar $2H$-$TaS_2$ system, wherein the suppression is derived by the pressure[31].

Next, we examined the strength of each coupling upon intercalation. First, the total coupling strength $\lambda_{tot}$ was estimated from the slope of Re$\Sigma$ at $E_F$, by definition. A monotonic increase of $\lambda_{tot}$ was revealed with increasing intercalation level (Figure 4e). This tendency is consistent with that extracted from the transport property, which accounts the enhancement of superconducting transition temperature $T_c$[2, 3]. As two different couplings contribute to $\lambda_{tot}$, the strength of each coupling was separated by considering the slope of Re$\Sigma$ at higher binding energy levels between the e-amp and e-ph kinks, as shown in the inset of Figure 4e and Figure S11. The resulting separation in Figure 4f reveals a different trend between $\lambda_{e\text{-amp}}$ and $\lambda_{e\text{-ph}}$. Interestingly, only $\lambda_{e\text{-amp}}$ increases while $\lambda_{e\text{-ph}}$ remains almost constant, indicating that $\lambda_{tot}$ enhancement is dominated by $\lambda_{e\text{-amp}}$. In the beginning, $\lambda_{e\text{-amp}}$ is comparable to $\lambda_{e\text{-ph}}$, at a value near 0.4. Over time, $\lambda_{e\text{-amp}}$ approaches a value ranging from 0.9–1.0, more than twice that of $\lambda_{e\text{-ph}}$, at the highest level.

The increase in $\lambda_{tot}$, primarily attributed to $\lambda_{e\text{-amp}}$, leads to the fascinating conjecture that e-amp coupling plays a specific role in $T_c$ enhancement. Figure 4g shows $T_c$ estimated by incorporating the collected coupling information into McMillan's equation[32] $T_c = \frac{\Theta_D}{1.45}\exp\left(-\frac{1.04(1+\lambda)}{\lambda-\mu^*(1+0.62\lambda)}\right)$. The average value of the two coupling energy scales at each intercalation level was substituted as the Debye temperature, along with $\lambda_{tot}$ at each level. The Coulomb pseudopotential $\mu^*$ was assumed as 0.25 and remained fixed for every level as the electron density is expected to maintain upon Pd intercalation. The estimated $T_c$ value agreed well with the $T_c$ value retrieved from transport measurements[2, 3]. This consistency ensures that the couplings included in $\lambda_{tot}$ constitute the source of superconductivity—the pairing mediator. Furthermore, the dominance of $\lambda_{e-amp}$ suggests that e-amp coupling is essential between these two couplings. We note that under e-ph coupling only, an extremely low $T_c$ value is expected so that $T_c$ enhancement upon Pd intercalation cannot be explained.

## 3. Discussion

The preceding results—the discovery of the signature of e-amp coupling and corresponding evolution towards the QCP—highlight the unprecedented role of CDW fluctuation in superconductivity formation, pairing electrons by amplitudons. Yet, the e-amp coupling has been traced only at the limited momentum positions, due to the previously mentioned complex gap opening. Still, the results could be representative for the possible coupling at other positions and for the formation of superconductivity, since expected optical-phonon-like flat amplitudon dispersion suggests a weak momentum dependence of e-amp coupling. Further, the known *s*-wave symmetry of superconductivity entails rather isotropic interaction source for the superconductivity[33-35].

Not only the possible role of e-amp coupling on the superconductivity, our results suggest a way to parametrically describe competing behavior. Our results reveal two trends of e-amp coupling as the system approaches the QCP of CDW. One trend entails amplitudon energy softening, and the other trend involves e-amp coupling strengthening. According to McMillan's equation, a reduction in amplitudon energy linearly reduces $T_c$ since the coupling energy determines the Debye temperature. Stronger e-amp coupling exponentially enhances $T_c$, as the coupling strength functions as an exponent. Therefore, in a particular well-balanced region, an increase in coupling strength can overcome amplitudon energy reduction and produce a higher $T_c$ value, which will eventually decrease to zero as the amplitudon energy becomes infinitesimal. Therefore, the dome-shaped superconducting phase near the QCP of the CDW can be regarded as a balance between these two factors.

The nature of these two trends must be elucidated. In the case of amplitudon energy reduction, e-amp kinks persist up to the coexisting phase where CCDW supposedly establishes a short-range order only. This indicates that the amplitudon can be excited from a short-range order or even with a certain degree of instability, similar to the case of the magnon observed without long-range spin ordering[36, 37]. However, in this case, a lower energy scale is expected than that of the excitation out of the well-stabilized long-range order, as mentioned above. Next, regarding the coupling strength, the Eliashberg-McMillan theory gives the dimensionless coupling constant as $\lambda = \int \frac{d\omega \alpha^2(\omega) F(\omega)}{\omega} = \frac{N(0)\langle g^2 \rangle}{M\langle \omega^2 \rangle}$ where $N(0)$ is the electronic density of states (DOS) at $E_F$, $g$ is the electronic matrix element, and $\omega$ is the phonon frequency[32]. According to the above definition, it can be easily expected that the suppression of CDW order (and so does CDW gap) recovers DOS at the Fermi surface $N(0)$, leading to the coupling strength enhancement. Also, the softening of the amplitudon energy, which reduces $\langle \omega^2 \rangle$, can strengthen the coupling. Therefore, the enhancement of coupling strength would be a natural consequence of the CDW suppression. Still, further theoretical and experimental investigation is needed to identify the actual mechanism of the enhancement, including other possible contributions such as the spectral weight of the amplitudon and van Hove singularity[3].

This compelling scenario of superconductivity by e-amp coupling requires further inquiries for substantiation. Future research should also address related questions, such as the way in which an electron can interact with self-governed order fluctuations, handling multiple couplings of different origins in electron pairing, the case of other competing orders in different systems, etc. We believe that these efforts will establish the microscopic mechanism behind the competing behavior, thereby elucidating the nature of the superconducting mechanism. The present results lay the foundation for such effort and, further, would be also informative in establishing the position of the CDW in cuprate superconductors and the connection between the non-trivial nature of both CDW and superconductivity in recent Kagome systems.

<Figure legends>

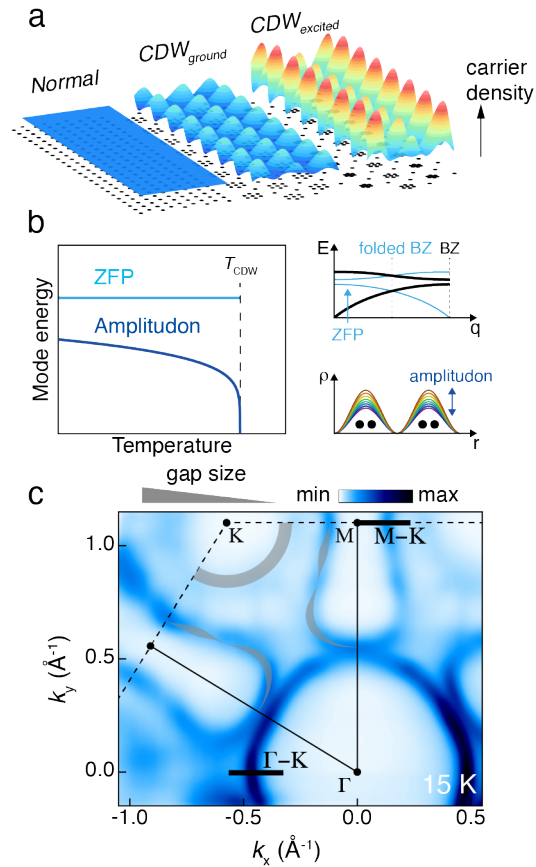

**Figure 1| Schematics of CDW-related excitations and the Fermi surface of 2*H*-TaSe$_2$.** a) Schematics of normal, CDW ground, and CDW excited states. b) Schematics of the CDW-related mode energy versus the temperature. The CDW amplitudon tends to soften with increasing temperature, while the ZFP energy remains constant. c) Fermi surface of 2*H*-TaSe$_2$ at 15 K, and CDW-gap-opened region in the Γ-K and M-K lines (the gray area; the width indicates the relative CDW gap size). The regions of interest, the Γ-K and M-K high-symmetry lines, are indicated with black bars.

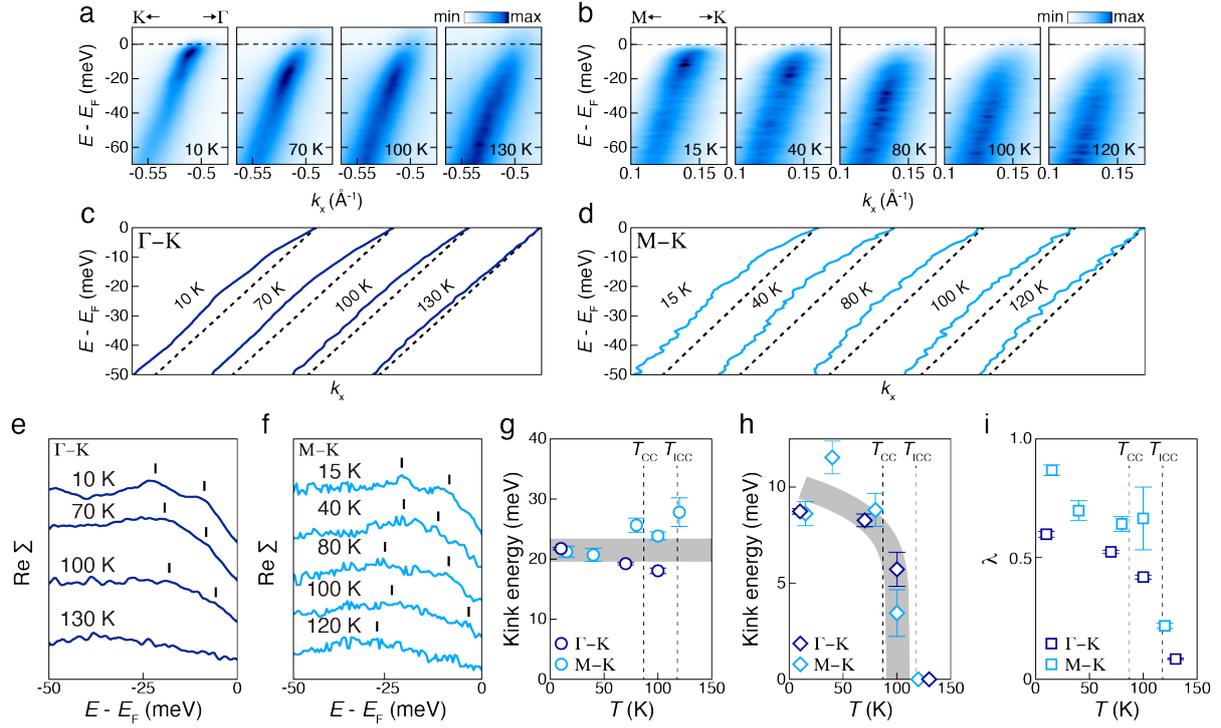

**Figure 2| Temperature-dependent kinks in 2H-TaSe$_2$.** a, b) Temperature-dependent Γ-K (a) and M-K (b) high-symmetry cuts. c, d) The peak positions obtained via MDC fitting of the Γ-K (c) and M-K (d) cuts. e, f) Temperature-dependent real part of the self-energy with offsets obtained by subtracting the estimated bare band dispersions (the dashed lines in (c) and (d)) from the peak positions (the colored solid lines in (c) and (d)) along the Γ-K (e) and M-K (f) lines. g, h) Higher (g) and lower (h) kink energy levels obtained from fitting to a continuous function consists of several linear lines (the black bars in (e) and (f)). The gray line in (g) is the average kink energy level at the lowest temperature while that in (h) is the data fitted to the softening function. (i) Coupling constant obtained from the slope of ReΣ at $E_F$.

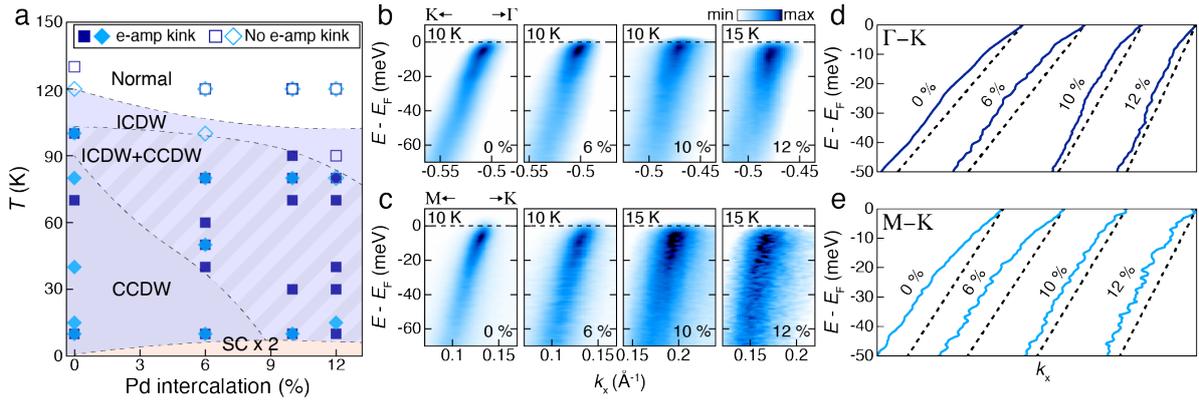

**Figure 3| Phase diagram and kinks of 2H-Pd$_x$TaSe$_2$.** a) Estimated phase diagram of 2H-Pd$_x$TaSe$_2$. The dark blue filled (empty) squares indicate where e-amp kinks are (not) detected in the Γ-K line, while the light blue diamond symbols are for the M-K line. b, c) Intercalation-dependent Γ-K (b) and M-K (c) high symmetry cuts. d, e) The peak positions of the Γ-K (d) and M-K (e) line obtained through MDC fitting.

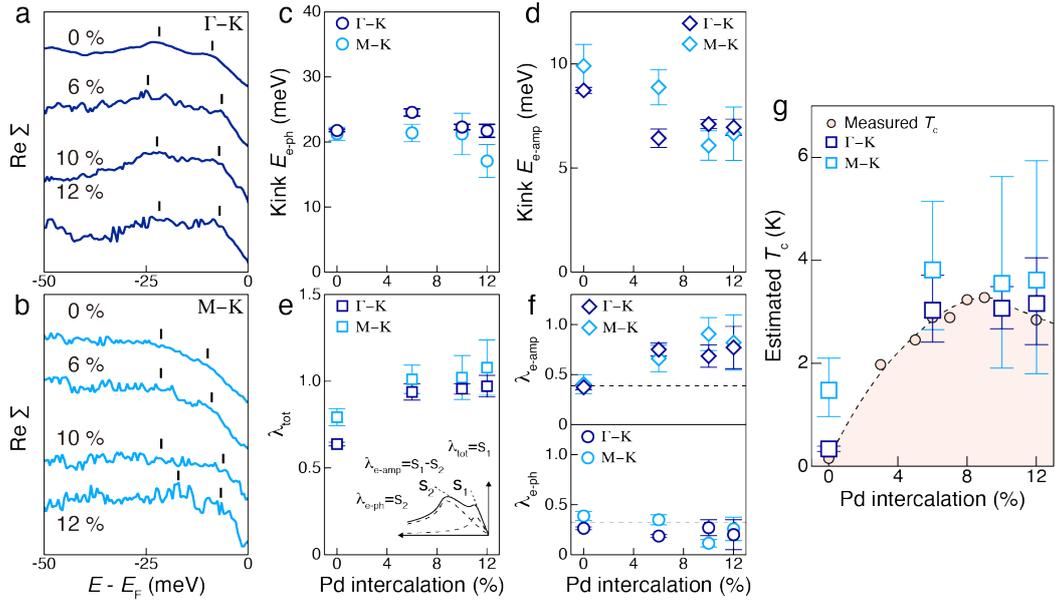

**Figure 4| Intercalation-dependent kink analysis.** a, b) The real part of the self-energy with offsets for Γ-K (a) and M-K (b) lines. c, d) Intercalation dependence of the e-ph (c) and e-amp (d) kink energies which are indicated as black bars in (a) and (b). e, f) Total, e-amp, and e-ph coupling constant as a function of intercalation level. The inset in (e) shows how each coupling constant is extracted from the real part of the self-energy. g) Estimated $T_c$ from the McMillan's equation with obtained parameters. The filled circles are the $T_c$ obtained by the transport measurement in ref. [2], and the dashed line is a guide for the eye.

<Method>

Single crystal of the pristine and Pd-intercalated 2$H$-Pd$_x$TaSe$_2$ samples were grown by chemical vapor transport (CVT) method. The transport agent of the pristine samples was I$_2$, while that of the intercalated samples was SeCl$_4$. In the case of the intercalated 2$H$-Pd$_x$TaSe$_2$, solid state reaction method was used before the CVT to synthesize the polycrystalline samples with different x values. The details of the 2$H$-Pd$_x$TaSe$_2$ sample growth and the characterization are described in ref. [2].

High-resolution ARPES was conducted with a synchrotron-radiation light sources of beamline 5-2 at the Stanford Synchrotron Radiation Lightsource (SSRL) of the Stanford Linear Accelerator Center and of beamline 5U at the Ultra Violet Synchrotron Orbital Radiation (UVSOR) of the Institute of Molecular Science. Supplementary high-resolution ARPES spectrum was obtained at the Korea Research Institute of Standards and Science (KRISS), with a He lamp as a light source. Time-of-flight ARPES was conducted with a laser light source at the Center for Correlated Electron Systems (CCES) of the Institute of Basic Science. A SCIENTA DA30 electron analyzer was used at SSRL to detect photoelectrons. The utilized incident photon energies for the main data were 48, 30 eV. The samples were cleaved in situ at temperature lower than 15 K under ultrahigh vacuum better than $1 \times 10^{-10}$ Torr. The energy resolution was 4 meV for 30 eV and 10 meV for 48 eV, estimated with the reference gold spectrum.

<Data availability>

The source data for Figure 1–4 is provided with the paper. All other data supporting the findings of this study are available from the corresponding author upon reasonable request.

<Acknowledgements>

We acknowledge helpful discussions with Y. Bang, H.-Y. Choi and E.-G. Moon. This research was supported by National R&D Program (No.2020K1A3A7A09080366), Creative Materials Discovery Program (No.2015M3D1A1070672), and Basic Science Resource Program (No.2020R1A4A2002828, No.2018R1D1A1B07050869) through the National Research Foundation of Korea (NRF) funded by the Ministry of Science and ICT. The use of the Stanford Synchrotron Radiation Lightsource, SLAC National Accelerator Laboratory, is supported by the U.S. Department of Energy, Office of Science, Office of Basic Energy Sciences under Contract No. DE-AC02-76SF00515. The work at SNU was financially supported by the National Research Foundation of S. Korea through 2019R1A2C2090648, 2019M3E4A1080227, and 2021R1A6C101B418. This research was supported by the KRISS (Korea Research Institute of Standards and Science) MPI (Materials Parts Instruments) Lab. program. The work at Ajou University was supported by the NRF funded by the Ministry of Education (No.2021R1A6A1A10044950, No.RS-2023-00285390). The reproduction of the data has been performed using facilities at IBS Center for Correlated Electron Systems, Seoul National University, and UVSOR Synchrotron Facility in the Institute of Molecular Science.

<Author contributions>

Y.K.K. conceived the work. Y.L., S.K., J.C., J.H., C.L., and Y.K.K. performed the ARPES measurements with support from M.H. and D.L. at SSRL, from S.I. and K.T. at UVSOR, and from Y.S.K., S.H., C.K. at IBS CCES. J.C. and S.H. grew the 2$H$-TaSe$_2$ single crystals, and Y.S. and K.H.K. grew the 2$H$-Pd$_x$TaSe$_2$ single crystals. Y.L. and Y.K.K. analyzed the ARPES data and all authors discussed the results. Y.L. and Y.K.K. wrote the manuscript in consultation with all authors.

<Competing interests>

The authors declare no competing interests.

# Supporting Information

## Coupling between electrons and charge density wave fluctuation and its possible role in superconductivity

Yeonghoon Lee[1,2], Yeahan Sur[3], Sunghun Kim[1,4], Jaehun Cha[1], Jounghoon Hyun[1], Chan-young Lim[1], Makoto Hashimoto[5], Donghui Lu[5], Younsik Kim[6,7], Soonsang Huh[6,7], Changyoung Kim[6,7], Shinichiro Ideta[8,9], Kiyohisa Tanaka[8], Kee Hoon Kim[3,10], Yeongkwan Kim[1]*

[1]*Department of Physics, Korea Advanced Institute of Science and Technology, Daejeon 34141, Republic of Korea.*

[2]*Quantum Technology Institute, Korea Research Institute of Standards and Science, Daejeon 34113, Republic of Korea.*

[3]*Center for Novel States of Complex Materials Research, Department of Physics and Astronomy, Seoul National University, Seoul 08826, Republic of Korea.*

[4]*Department of Physics, Ajou University, Suwon 16499, Republic of Korea.*

[5]*Stanford Synchrotron Radiation Light Source, Stanford Linear Accelerator Center, Menlo Park, CA, 94025, USA.*

[6]*Center for Correlated Electron Systems, Institute for Basic Science, Seoul 08826, Republic of Korea.*

[7]*Department of Physics and Astronomy, Seoul National University, Seoul 08826, Republic of Korea.*

[8]*Ultra Violet Synchrotron Orbital Radiation, Institute for Molecular Science, Myodaiji, Okazaki 444-8585, Japan.*

[9]*Hiroshima Synchrotron Radiation Center, Hiroshima University, Higashi-Hiroshima 739-0046, Japan.*

[10]*Institute of Applied Physics, Seoul National University, Seoul 08826, Republic of Korea.*

*e-mail: yeongkwan@kaist.ac.kr

**Momentum distribution curve fitting method and results**

To determine the precise peak positions and widths from the ARPES spectra, we have fitted the momentum distribution curves (MDCs) for all data sets. For the case of M-K and Γ-M high symmetry lines of 2$H$-TaSe$_2$, there are multiple bands (four for M-K, and two for Γ-M) which can affect the fitting, so all corresponding peaks were included in the fit function. In Figure S1 – S4, MDC fitting results of the all datasets used in the main text are presented. It is clear that all the fit lines (black solid lines) closely reproduce the raw data (colored open circles).

**Estimation of bare band dispersion**

To estimate the bare band dispersion, the peak positions obtained by MDC fitting are fitted with a polynomial function. Then, the bare band dispersion is determined as a linear line connecting two peak positions, one at $k_F$ and the other at the high binding energy of the highest temperature. In Figure S5, the extended peak positions obtained by MDC fitting and the estimated bare band is plotted.

**Fermi surface of Pd-intercalated 2$H$-TaSe$_2$**

In Figure S6, the Fermi surfaces of each 2$H$-Pd$_x$TaSe$_2$ (x=0.06, 0.10 and 0.12) are displayed. As expected, the band folding feature and the CDW gap opening were disappeared at high intercalation level. It demonstrates that the long-range CDW order is completely suppressed at intercalation level higher than 10 %.

**Temperature- and intercalation-dependent kink**

The temperature and intercalation dependence of kinks are presented with the real part of self-energies (ReΣ), given in Figure S7. The same analysis procedure used for Figure 2 in the main text was applied for all cases. For both Γ-K and M-K cases, the line shape analysis of peak position evidently indicates the existence of temperature-dependent kink. ReΣ can be extracted as mentioned in the main text, and from those, kink energies were estimated. Particularly, in the case of low intercalation level, the softening behavior for the e-amp kink upon increasing temperature is well reproduced.

**Simulation on the thermal broadening effect**

To check whether thermal broadening effect interferes the detection of the kink with small energy scale of around 10 meV at high temperature, the simulation for the temperature-dependent kink was performed. The bare band was assumed to be linear and fixed. The energy resolution effect is introduced by convoluting the generated spectra with Gaussian function of width 10 meV, which is the largest value for the main data. The energy scale of the kink is set to 10 meV for all temperature to solely examine the thermal broadening effect. As given in Figure S8, a kink is apparent even in the high-temperature case. Further, a kink can be more clearly revealed by the fitting the simulated spectra with MDC. Based on the simulation, it can be concluded that the kink structure cannot be erased by the thermal broadening effect.

**Temperature dependence of the electron-phonon coupling**

To estimate the temperature dependence of band renormalization produced by electron-phonon coupling, we calculated the renormalization of band including the temperature dependence of the self-energy. Normally, there are three main contributions to the self-energy: $\Sigma''_{e-ph}$ (electron-phonon scattering), $\Sigma''_{e-e}$ (electron-electron scattering), and $\Sigma''_{e-df}$ (electron-defect interactions) [1-3]. It is known that $\Sigma''_{e-df}$ is not usually strong and serves as a constant offset, and the effect of $\Sigma''_{e-e}$ is small near the Fermi level and thus its temperature dependence. Therefore, only $\Sigma''_{e-ph}$ with offset was accounted for the temperature dependence, which has the form of

$$\left|\Sigma''_{e-ph}\right| = \pi\hbar \int_0^{\omega_{max}} \alpha^2 F(\omega')[1 - f(\omega - \omega') + 2n(\omega') + f(\omega + \omega')]d\omega' \quad (S1)$$

where $\alpha^2 F(\omega)$ is the Eliashberg coupling function, $f$ is the Fermi-Dirac distribution function, and $n$ is the Bose-Einstein distribution function. $\alpha^2 F(\omega)$ is assumed to have the shape of a Debye spectrum,

$$\alpha^2 F(\omega) = \begin{cases} 3A|\omega|^2/\omega_D^3, & \text{if } |\omega| < \omega_D \\ 0, & \text{if } |\omega| \geq \omega_D \end{cases} \quad (S2)$$

and the real part of the self-energy $\Sigma'$ is retrieved using Kramers-Kronig relations.

For the kink energy, the average kink energy at the M-K line is used for the simulation, which is 24.5 meV. The constant $A$ of the equation (S2) is roughly determined to $5 \times 10^{12}$ which reproduces well the actual M-K data. Note that the exact value for the constant $A$ is not necessary as the overall trend of the coupling constant is not affected by it. We used the estimated bare band of M-K in Figure S5e as a bare band. Inserting all, the renormalization of band dispersion at various temperature is simulated (Figure S9). As the temperature is raised, the kink structure gradually broadens but does not disappear. The solid curve of coupling strength in Fig. S9b shows that the coupling strength reduces only by a half even at 150 K thus the kink should be visible.

**Ruling out optical and acoustic phonons as an origin of the low-energy kink**

It is easy to show that optical phonon is not the one that induces the low-energy kink as there is no optical phonon with energy less than 15 meV [4-8]. On the other hand, the case of acoustic phonon requires delicacy. The energy level of the acoustic phonon matches with the low-energy kink, and recent inelastic x-ray scattering study on 2$H$-TaSe$_2$ have shown that acoustic phonon also softens toward $T_{CDW}$ when the wave vector $q$ is close to $q_{CDW}$, although it recovers its energy above $T_{CDW}$ [9]. This makes it hard to exclude the possibility that e-acoustic-phonon coupling induces the low-energy kink. To rule out such possibility, additional high-resolution ARPES mapping was conducted (Fig. S10). To compensate the low statistics, which is a tradeoff for high resolution, denoising technique [10] is used to get clean spectrum (Fig. S10b). The band dispersions from MDC fitting and the resulting kink energies in Fig. S10c and S10d shows that the kink energies are almost constant upon momentum variation. If the low-energy kink is indeed from the acoustic phonon, the kink energy should soften to zero when the nesting condition is met [9, 11]. As no such anomaly is seen at any momentum, we conclude that the acoustic phonon is not the source of the low-energy kink.

**Separation method of two adjacent kinks**

To separate two different kinks avoiding intentional choice, a continuous function consisting of several linear lines is used as a fitting function (Figure S11). The vertices $a_0$ and $a_2$ are defined as $E_{amp}$ and $E_{ph}$, and the slopes $a_4$ and $a_6$ are defined as $\lambda_{tot}$ and $\lambda_{e-ph}$. The e-amplitudon coupling constant $\lambda_{e-amp}$ is calculated by subtracting $\lambda_{tot}$ and $\lambda_{e-ph}$.

To verify whether the coupling constants are properly separated, the Debye model is used to fit the real part of the self-energy $\Sigma'$. From equation (S1) and (S2), the imaginary part of the self-energy $\Sigma''$ from the e-phonon coupling can be reduced to,

$$|\Sigma''_{e-ph}(\omega)| = \begin{cases} \hbar\lambda\pi|\omega|^3/(3\omega_D^2), & \text{if } |\omega| < \omega_D \\ \hbar\lambda\pi\omega_D/3, & \text{if } |\omega| \geq \omega_D \end{cases}, \quad (S3)$$

at sufficiently low temperature [1,2]. With additional polynomial function $\beta x^2$ from $\Sigma''_{e-e}$ contribution, one can calculate the real part of the self-energy using the Kramers-Kronig transformation. Using this as a fitting function, the coupling strengths of the two kinks are separated in Figure S12. The resulting trend of the coupling strengths is consistent with our analysis; $\lambda_{\text{tot}}$ and $\lambda_{\text{e-amp}}$ significantly increases upon Pd intercalation while $\lambda_{\text{e-ph}}$ does not. This result implies that the definition in Figure S11 is appropriate.

**The effect of band hybridization on coupling constant analysis**

Distinct feature of the real part of the self-energy of 2$H$-Pd$_x$TaSe$_2$ is that the self-energy does not converge to zero even at high binding energy regime. Indeed, the band dispersion at the lowest temperature does not meet the estimated bare band dispersion (Figure S13). The most probable explanation of this uplift is a gap induced by band hybridization. When the system is in the CDW state, the hybridization of the main band and the folded band induces a gap in the electronic structure (Figure S14). By using a simple model of gapped dispersion relation [12],

$$E_{\pm}(k) = \frac{\hbar^2}{2m}\left(\frac{1}{4}G^2 + \left(k - \frac{1}{2}G\right)^2\right) \pm \sqrt{4\left(\frac{\hbar^2}{2m}\left(\frac{1}{2}G\right)^2\right)\left(\frac{\hbar^2}{2m}\left(k - \frac{1}{2}G\right)^2\right) + U^2}, \quad (S4)$$

it is possible to simulate the band dispersion of 2$H$-TaSe$_2$ in the CDW state (Figure S14a and b). Although this hybridization considerably renormalizes the band dispersion, its effect on coupling constant is negligible due to the parabolic feature of the renormalization (Figure S14f).

**Intercalation-dependent analysis of the imaginary part of the self-energy**

If we assume that the bare dispersion is almost linear near the Fermi level, *i.e.*, $\varepsilon(k) = vk$, we can calculate the imaginary part of self-energy $\Sigma''$ from MDC width $FWHM = 2|\Sigma''(\omega)/v|$. The results are plotted in Figure S15a (dark blue circles). And then, the imaginary part of the self-energy is fitted to equation (S3) with additional polynomial function $\beta x^2$ for $\Sigma''_{e-e}$ contribution [1], as the system shows Fermi liquid behavior below 30 K [13]. From the fitted parameters, it is possible to extract the total $\lambda_{\text{tot}}$, e-amp $\lambda_{\text{e-amp}}$, and e-ph coupling constants $\lambda_{\text{e-ph}}$ (Figure S15b and c). With those parameters, we also calculated $T_c$ separately to cross check those estimated with parameters extracted from the real part of self-energy (Figure S15d). Coupling constants were extracted with 2 different sets of imaginary parts of self-energy at each intercalation level and averaged. The tendency of the estimated $T_c$ fits with that obtained from the real part of self-energy. although slightly larger value for the Coulomb pseudopotential μ* was used compared to the real part of the self-energy analysis. This consistency between two independent analysis supports our interpretation that the coupling constant is enhanced due to the intercalation and the major cause of the enhancement is e-amp coupling.

It is also possible to check the validity of our definition of the real part of the self-energy $\Sigma'$ using the imaginary part $\Sigma''$. If $\Sigma'$ is properly extracted, $\Sigma'$ and $\Sigma''$ should be connected by the Kramers-Kronig transformation [1,2]. In Figure S16, $\Sigma''$ obtained by Kramers-Kronig transformation is compatible with $\Sigma''$ from the MDC width. This correspondence verifies that the real part of the self-energy is well-defined.

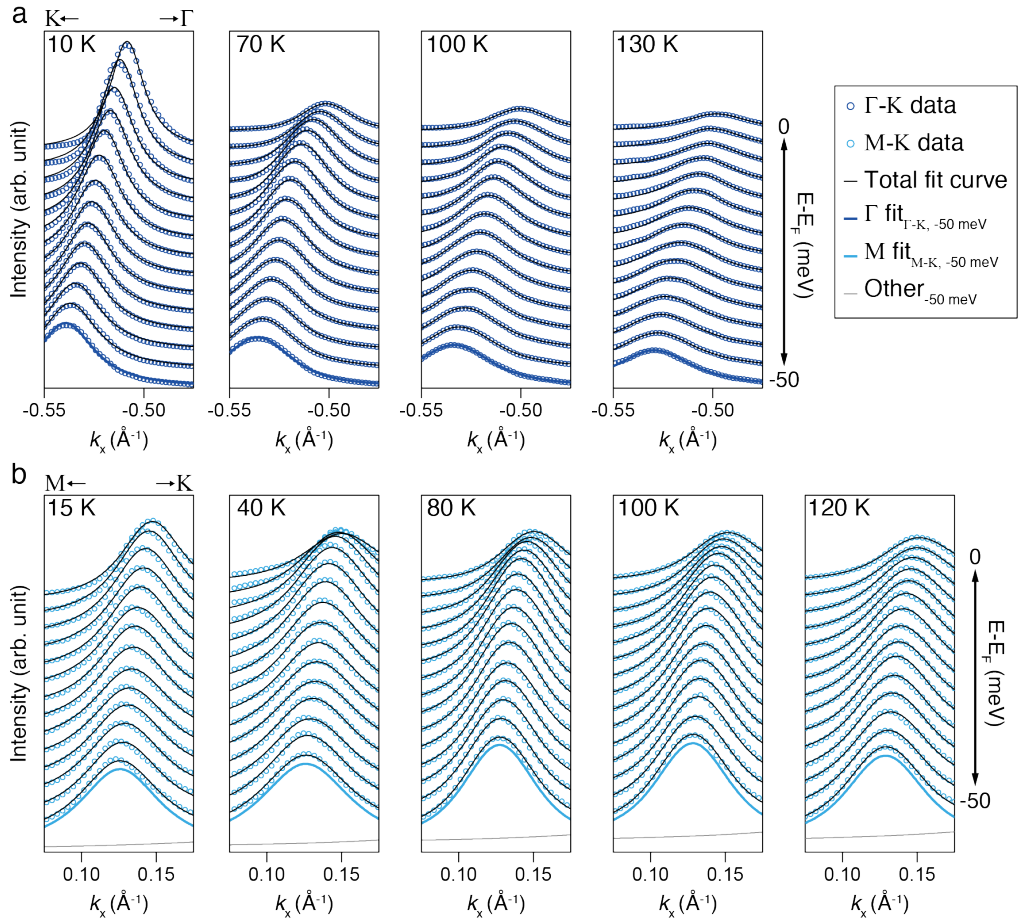

**Supplementary Figure S1.**

a, b) MDC fitting results of pristine TaSe$_2$ at the Γ-K (a) and M-K (b) high symmetry lines with various temperatures and binding energies. Dark/light blue empty circles are the raw data, black solid lines are total fit curve including all necessary terms. Dark/light blue solid lines are single-peak fittings for each spectrum, and gray lines in (b) are additional peak of K band right near the M band where the peak position locates outside of the plotted momentum range. The energy step of MDC stack is approximately 3.6 meV; some data points in both energy and momentum directions are not plotted for the better visualization.

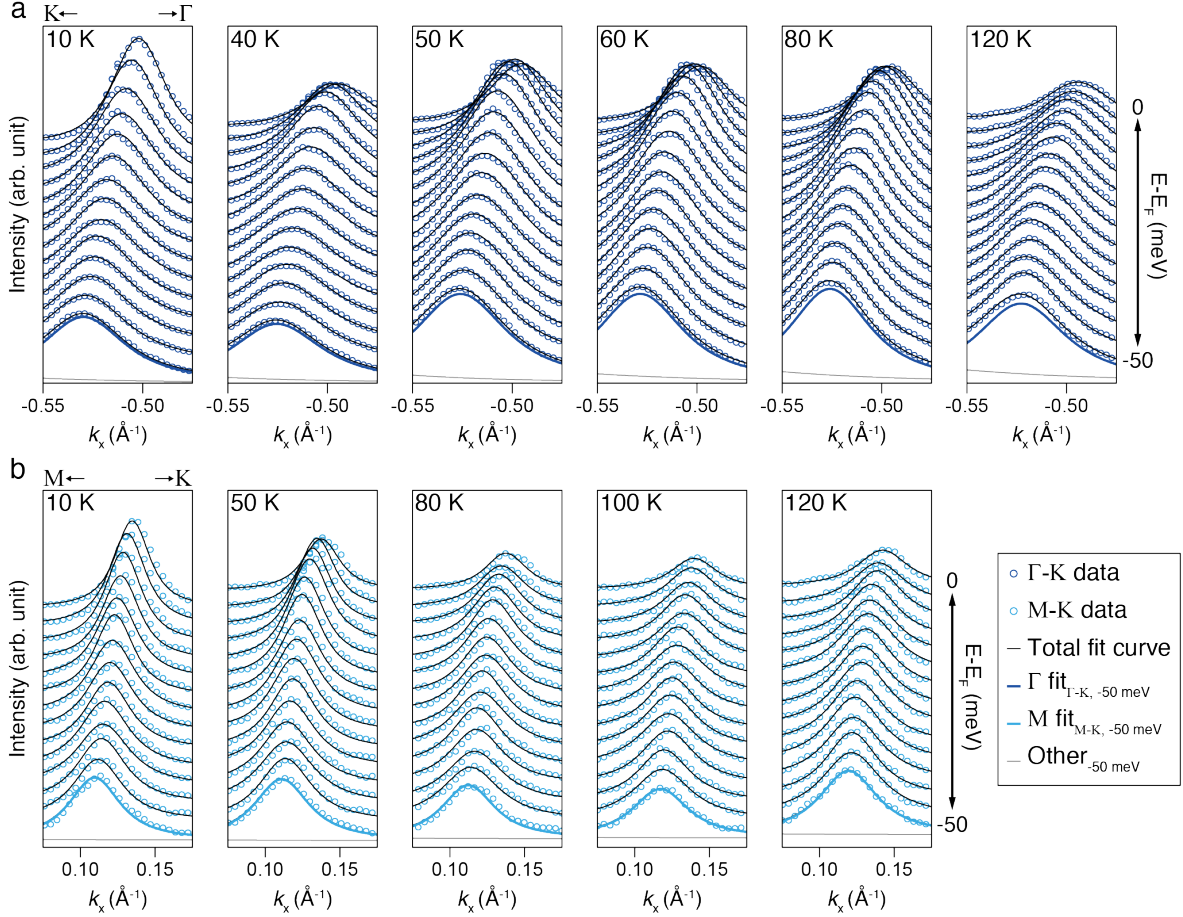

**Supplementary Figure S2.**

a, b) MDC fitting results of $Pd_{0.06}TaSe_2$ at the Γ-K (a) and M-K (b) high symmetry lines with various temperatures and binding energies. Dark/light blue empty circles are the raw data, black solid lines are total fit curve including all necessary terms. Dark/light blue solid lines are single-peak fittings for each spectrum, and gray lines are additional peaks where the peak positions locate outside of the plotted momentum range. The energy step of MDC stack is approximately 3.6 meV; some data points in both energy and momentum directions are not plotted for the better visualization.

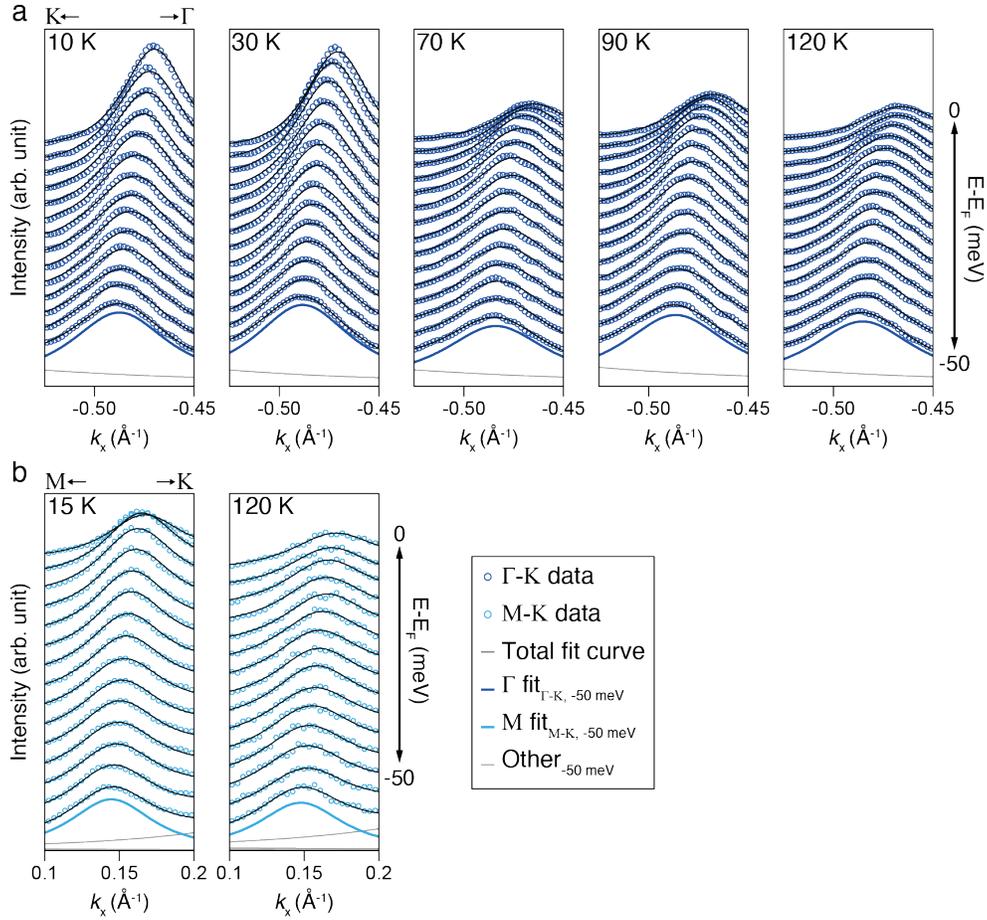

**Supplementary Figure S3.**

a, b) MDC fitting results of Pd$_{0.10}$TaSe$_2$ at the Γ-K (a) and M-K (b) high symmetry lines with various temperatures and binding energies. Dark/light blue empty circles are the raw data, black solid lines are total fit curve including all necessary terms. Dark/light blue solid lines are single-peak fittings for each spectrum, and gray lines are additional peaks where the peak positions locate outside of the plotted momentum range. The energy step of MDC stack is approximately 3.6 meV; some data points in both energy and momentum directions are not plotted for the better visualization.

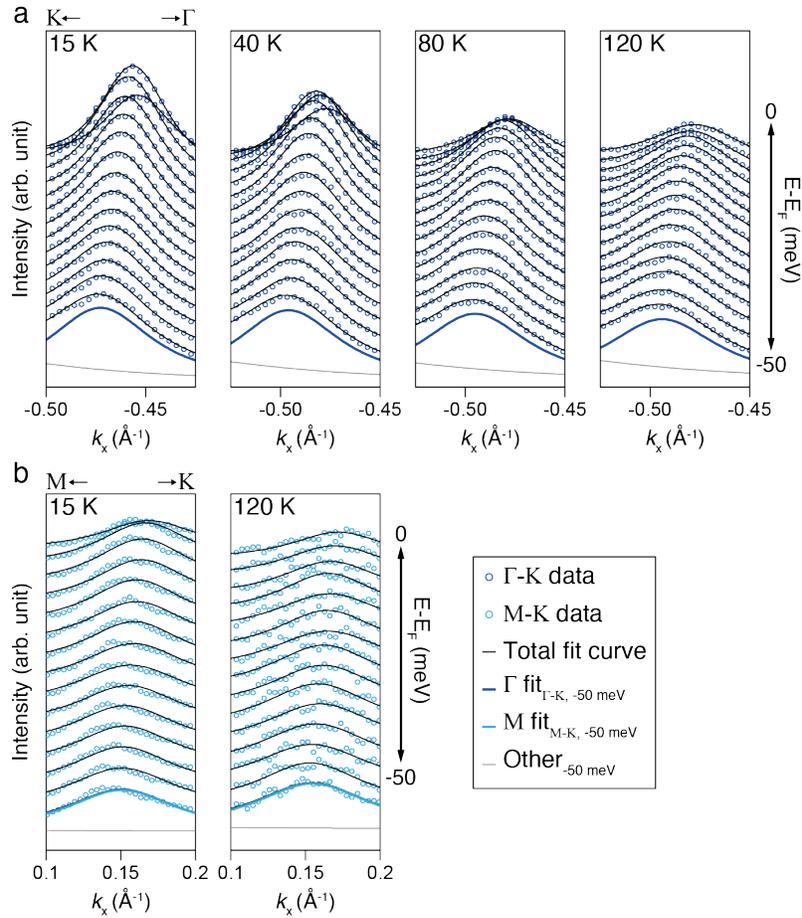

**Supplementary Figure S4.**

a, b) MDC fitting results of Pd$_{0.12}$TaSe$_2$ at the Γ-K (a) and M-K (b) high symmetry lines with various temperatures and binding energies. Dark/light blue empty circles are the raw data, black solid lines are total fit curve including all necessary terms. Dark/light blue solid lines are single-peak fittings for each spectrum, and gray lines are additional peaks where the peak positions locate outside of the plotted momentum range. The energy step of MDC stack is approximately 3.6 meV; some data points in both energy and momentum directions are not plotted for the better visualization.

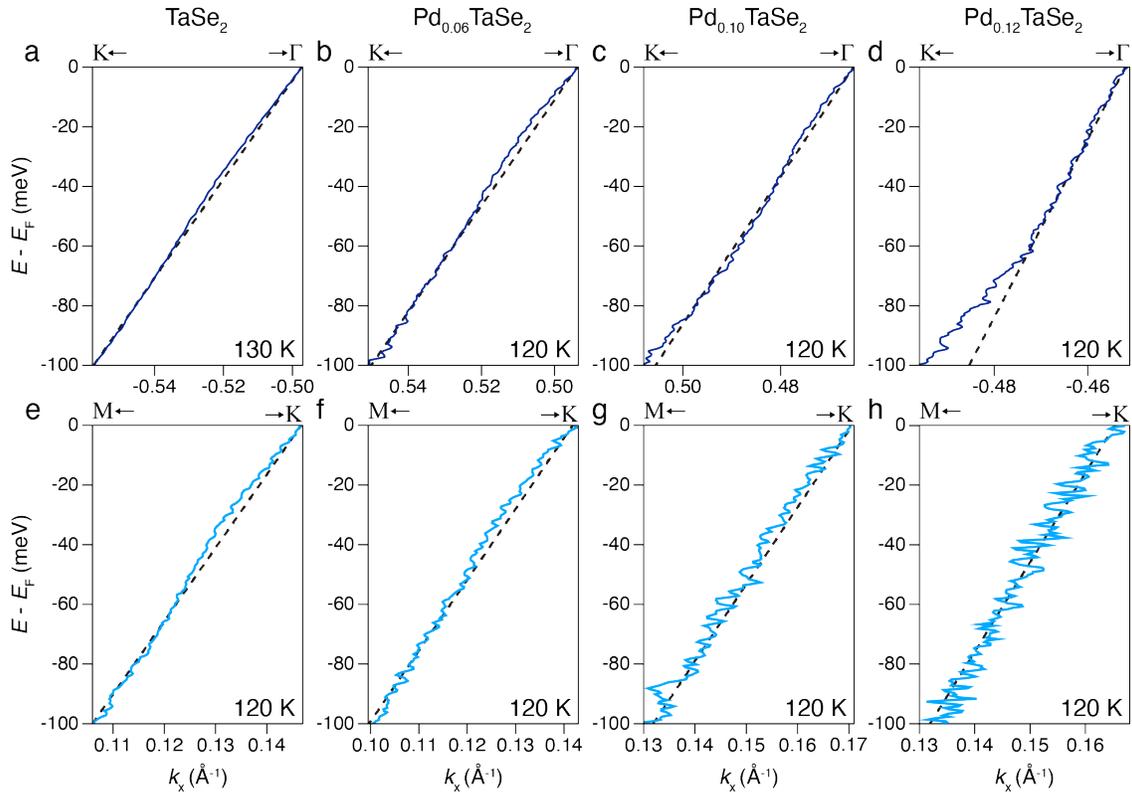

**Supplementary Figure S5.**

a–h) Peak positions for each intercalation level at the highest temperature obtained by MDC fitting (colored solid lines) and the estimated bare band (dashed lines) along the Γ-K (a to d) and M-K (e to h) high symmetry lines.

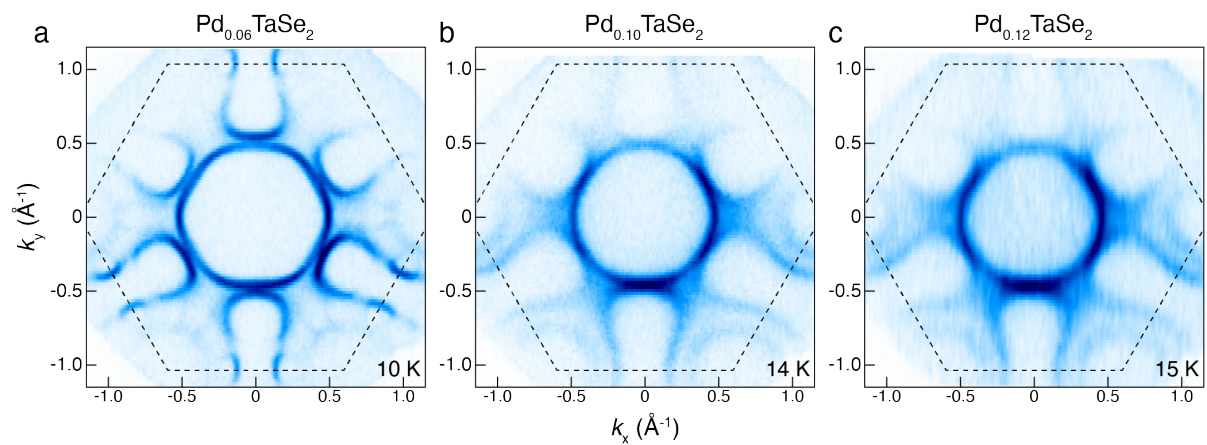

**Supplementary Figure S6.**
a–c) Fermi surfaces of 2$H$-Pd$_{0.06}$TaSe$_2$ (a), 2$H$-Pd$_{0.10}$TaSe$_2$ (b) and 2$H$-Pd$_{0.12}$TaSe$_2$ (c).

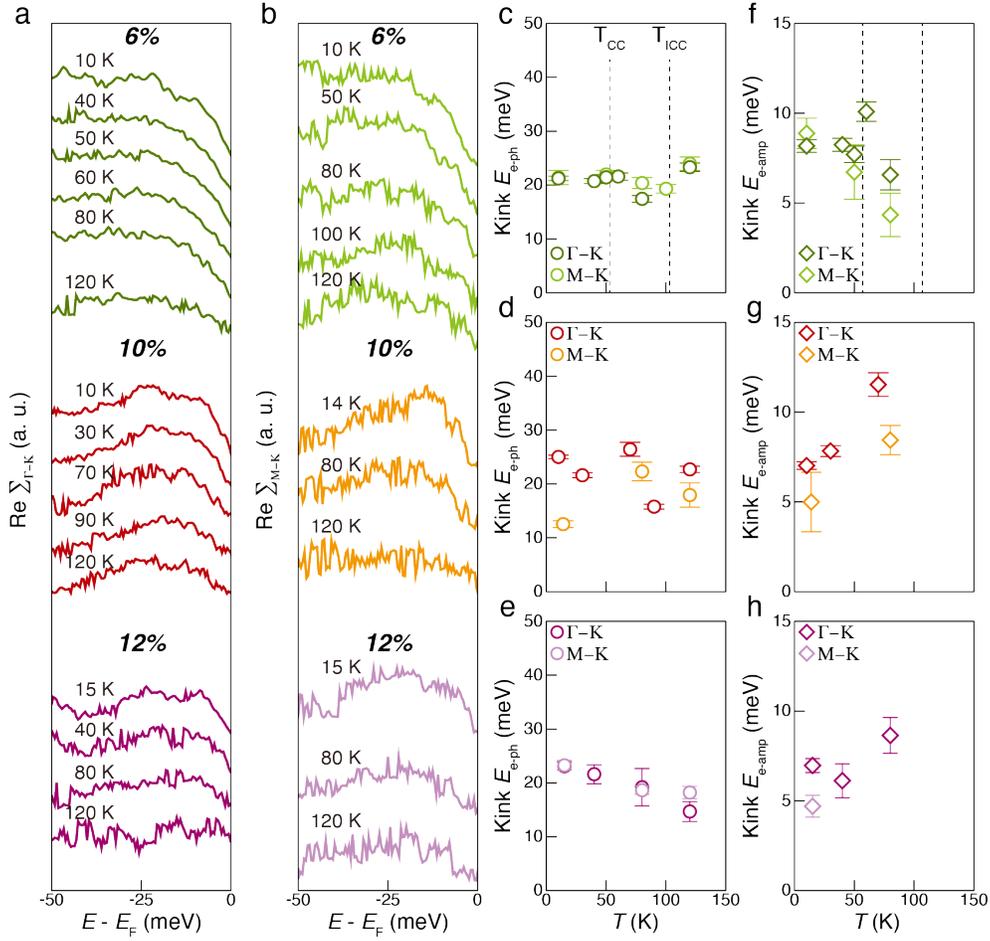

**Supplementary Figure S7.**

a, b) Temperature- and intercalation-dependent real part of the self-energies for Γ-K (a) and M-K (b), plotted with offsets. c–h) Temperature dependence of the e-ph (c-e) and e-amp (f-h) kink energies obtained from fitting to a continuous function consists of several linear lines (refer Figure S11).

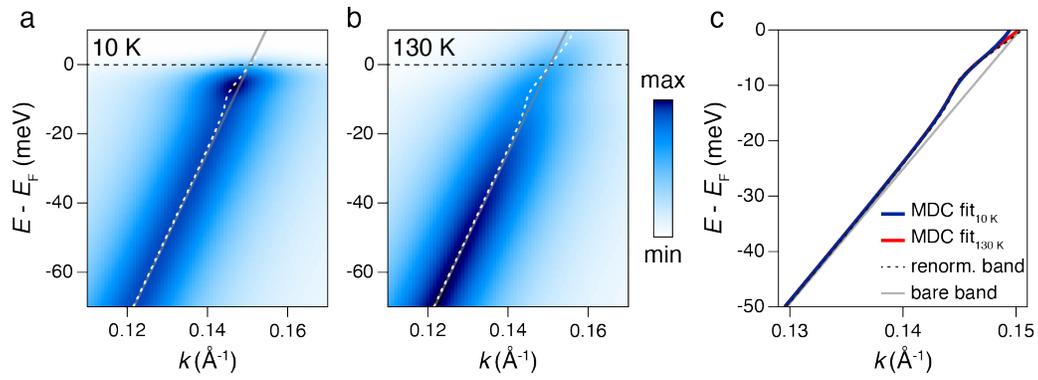

**Supplementary Figure S8.**

a, b) Simulated spectra for 10 K (a) and 130 K (b). Other than the thermal broadening effect, all the parameters are kept. The dashed lines are peak positions with kink (renormalized band dispersions), while the gray solid lines are bare band dispersions. c) Peak positions of (a) and (b) obtained by the MDC fitting of the simulated spectra.

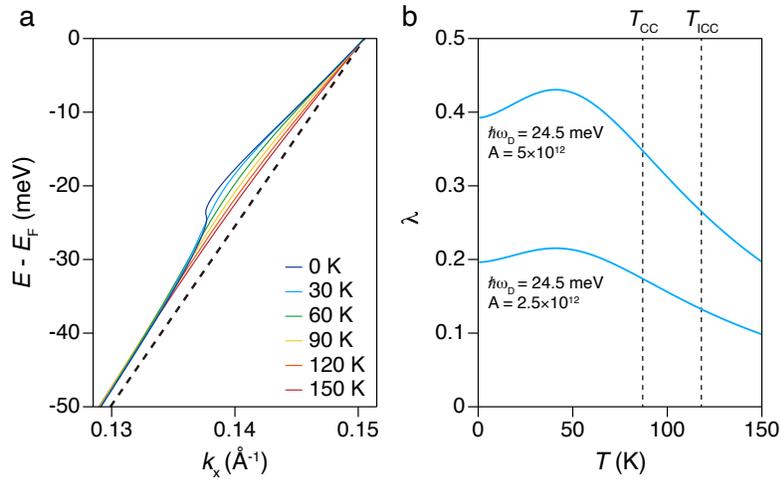

**Supplementary Figure S9.**
a) Simulated band dispersions with 24.5 meV kink at various temperatures. A Debye spectrum is used for the Eliashberg coupling function. b) Temperature dependent coupling constant of the simulated dispersions with different $A$ values, the weight constant in the Debye spectrum.

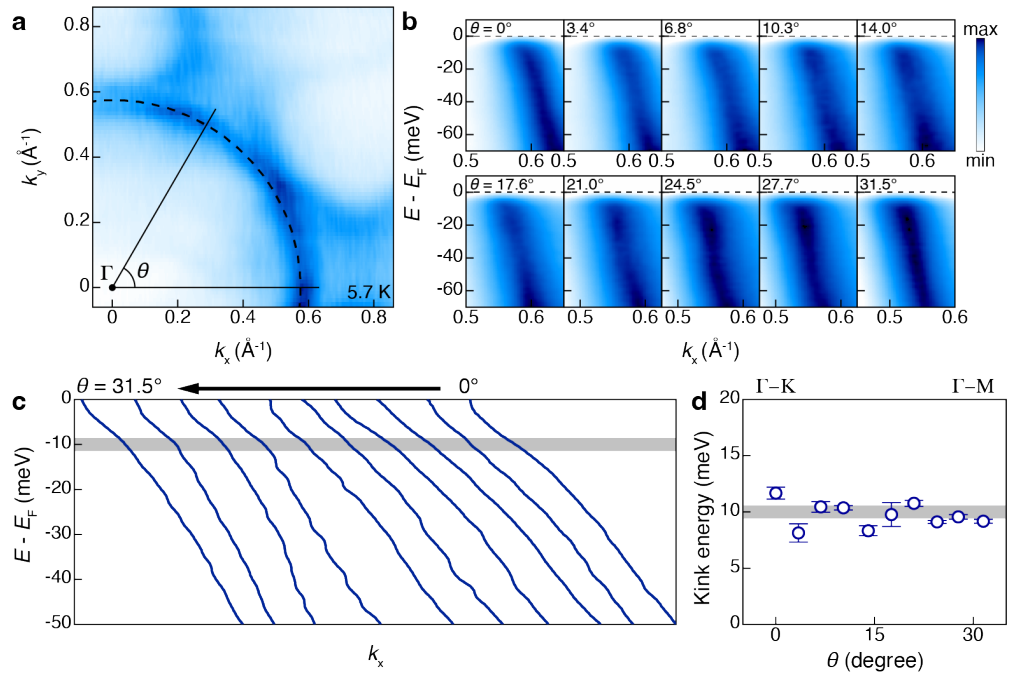

**Supplementary Figure S10.**

a) Fermi surface of 2$H$-TaSe$_2$ obtained by ARPES. Dashed line is a Γ-band guide for the eye. b-d) Momentum-dependent ARPES spectra (b), band dispersion extracted from the MDC fitting (c), and the lowest kink energy (d) of the Γ band.

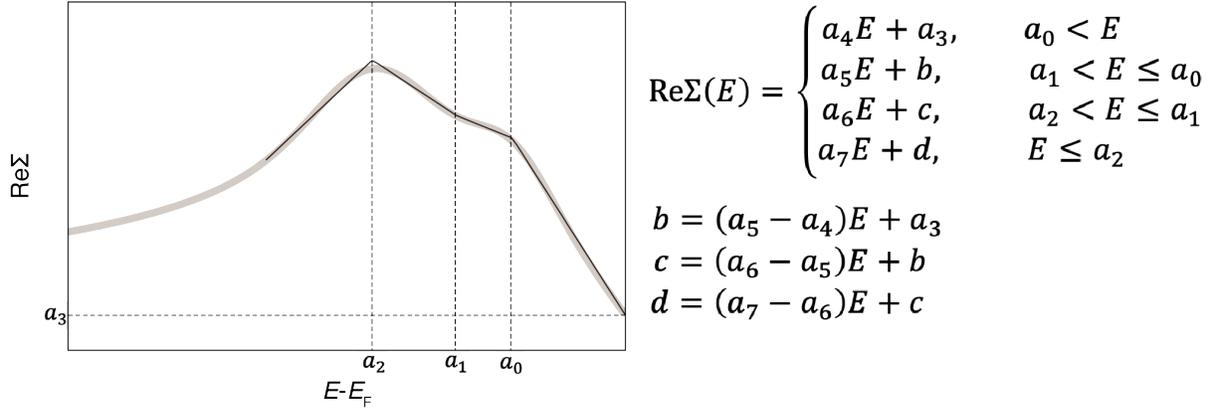

**Supplementary Fig. S11.**
The way to separate two kinks. The real part of the self-energy is fitted with a continuous function consisting of multiple linear lines. The vertices $a_0$ and $a_2$ are defined as kink energies, and the slopes $a_4$ and $a_6$ are used to calculate coupling constants.

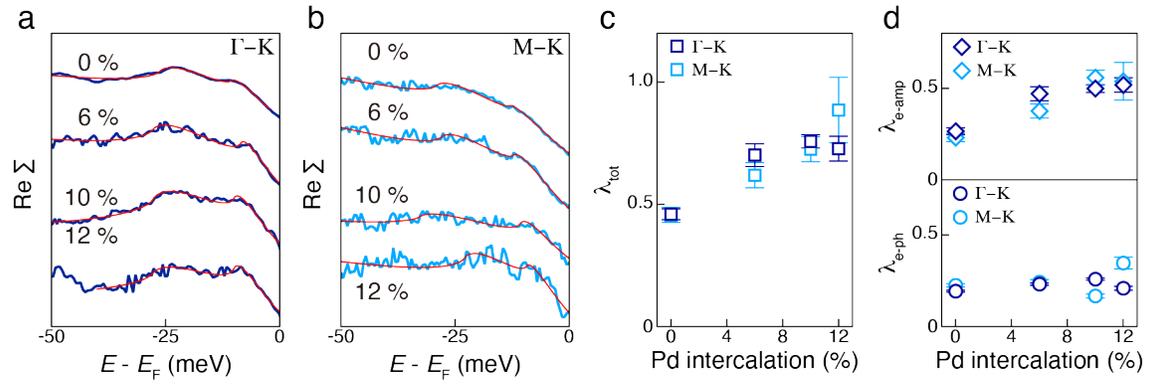

**Supplementary Figure S12.**
a, b) Intercalation dependence of the real part of the self-energies, fitted with the Debye model (red curves). c, d) Intercalation dependence of the $\lambda_{tot}$ (squares), $\lambda_{e-amp}$ (diamonds), and $\lambda_{e-ph}$ (circles).

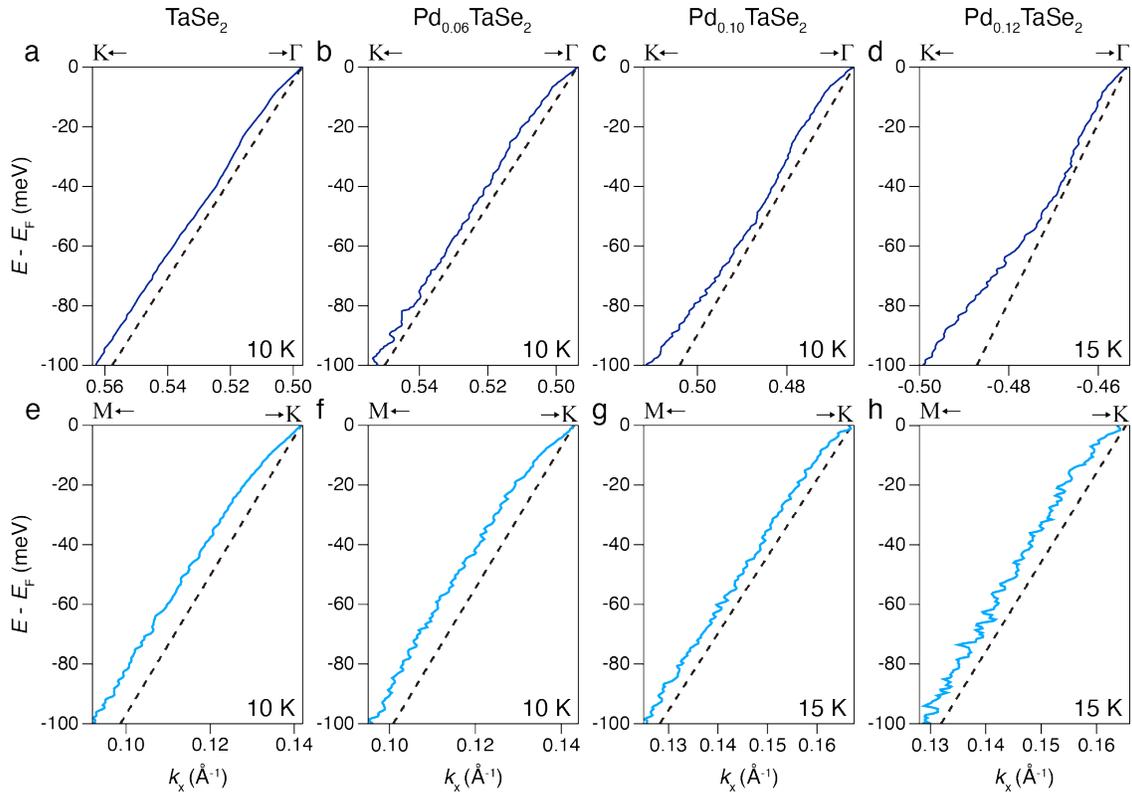

**Supplementary Figure S13.**

a-h) Peak positions for each intercalation level at the lowest temperature obtained by the MDC fitting (colored solid lines) and the estimated bare band (dashed lines) at the Γ-K (a-d) and the M-K (e-h) high symmetry lines.

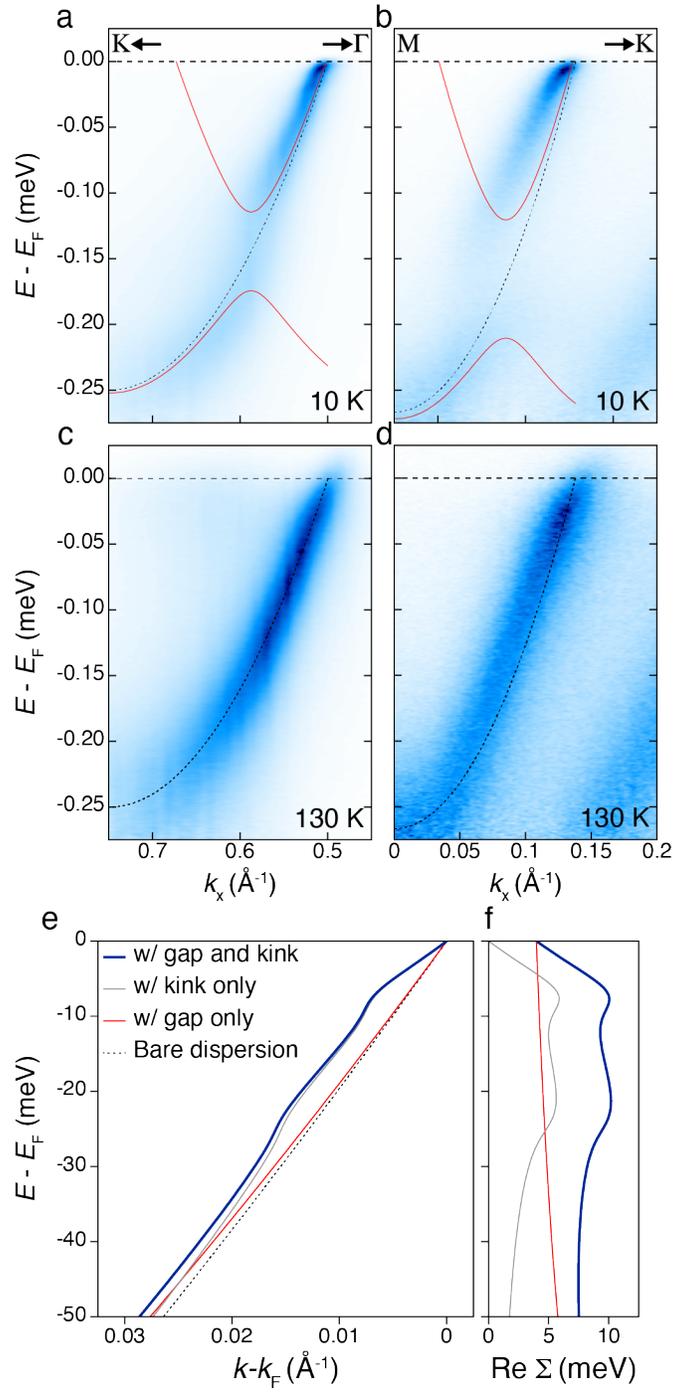

**Supplementary Figure S14.**

a-d) The ARPES spectra of 2$H$-TaSe$_2$ at the CCDW phase (a, b) and the normal phase (c, d) with band calculation. Dashed lines in (a) to (d) are the parabolic bare band dispersions, and red solid lines in (a) and (b) are the gapped dispersions. e, f) Magnified image of the calculated band dispersions (e) and the corresponding real part of the self-energy (f), with and without kinks and a hybridization gap.

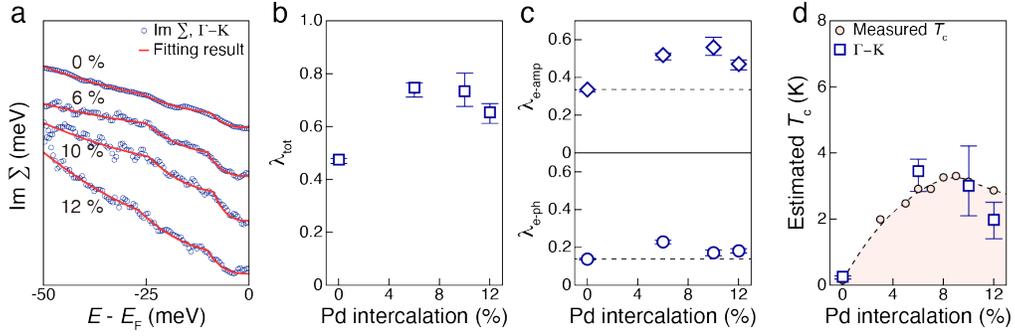

**Supplementary Figure S15.**

a) Intercalation dependence of the imaginary part of the self-energy (blue empty circles) and its fitting results (red solid lines). b, c) Intercalation dependence of total, e-amp, and e-ph coupling constant extracted from the fitting. d) Estimated $T_c$ with the McMillan's equation. Coulomb pseudopotential $\mu^*$ is assumed to be constant value 0.18. and the Debye temperature $\Theta_D$ is inserted with the average of the e-ph and e-amp kink energies. The filled circles are the $T_c$ obtained by the transport measurement in ref. [13] and the dashed line is a guide.

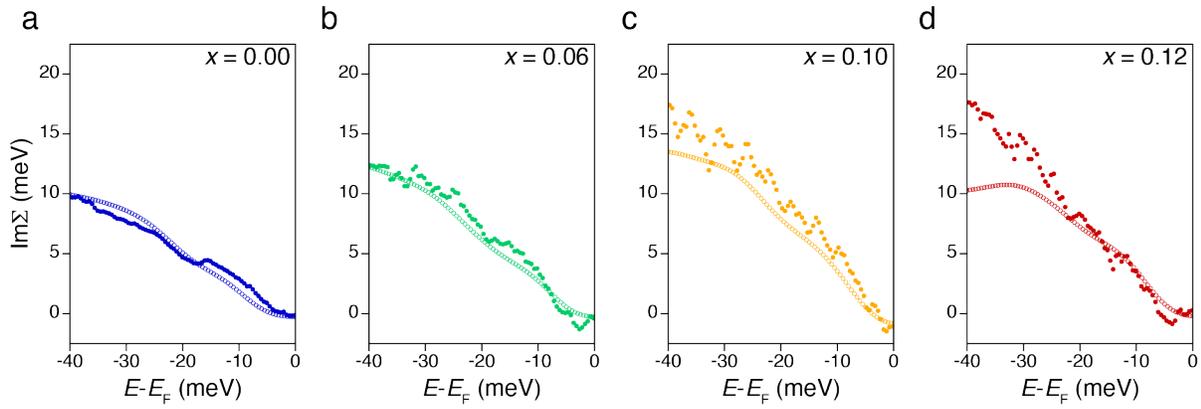

**Supplementary Figure S16.**
a-d) The imaginary part of the self-energies with offsets calculated by two different methods: MDC peak width multiplied by bare Fermi velocity (filled circles), and applying Kramers-Kronig transformation to the real part of the self-energy (empty circles).